\documentclass[12pt]{article}

\usepackage{graphicx}
\usepackage{dcolumn}
\usepackage{bm}
  \def\be{\begin{equation}}      \def\ol{\overline}
  \def\ee{\end{equation}}    \def\beq{\begin{eqnarray}}
  \def\dis{\displaystyle}    \def\eeq{\end{eqnarray}}
  \def\btd{\bigtriangledown}     \def\m{\multicolumn}
  \def\dfac{\dis\frac}       \def\ra{\rightarrow}

\begin{document}

\begin{center}
{\Large Properties of $Q\bar Q$ mesons in non-relativistic QCD formalism}\\

{\large Ajay Kumar Rai$^\dag$, Bhavin Patel$^*$ and P C Vinodkumar}\\

$^\dag$Physics Section, Applied Sciences and Humanities Department, Sardar Vallabh \\
National Institute of Technology, Surat-395 007, Gujarat, India.\\
$^*$Department of Physics, Sardar Patel University, Vallabh Vidyanagar,\\
Anand - 388 120, Gujarat,India.\\
Email:raiajayk@rediffmail.com
\end{center}
\begin{abstract}
The decay rates of $Q \bar Q$ mesons ($Q\ \varepsilon \ {c, b}$) are
studied in the NRQCD formalism in terms of their short distance and
long distance coefficients. The long distance coefficients are
obtained through phenomenological potential model description of the
mesons. The model parameters that reproduces the mass spectrum of
the $c \bar c$, $b \bar b$ and $c \bar b$ mesons are employed to
study the decay widths of these mesons. We extract the mass spectrum
as well as the reproduces the respective radial wave functions from
the different potential models as well as from non-relativistic
phenomenological quark antiquark potential of the type
$V(r)=-\frac{\alpha_c}{r}+A r^{\nu}$, with $\nu$ varying from 0.5 to
2. The spin hyperfine and spin-orbit interactions are employed to
obtain the masses of the pseudoscalar and vector mesons. The decay
constants with QCD corrections are computed in this model as well as
in the case of other potential models for comparison. The digamma
and dileptonic decays of $c \bar c$, and $b \bar b$ mesons are
investigated using some of the known potential models without and
with radiative corrections up to the lowest order. These decay width
are also computed within the NRQCD formalism up to $O(v^4)$ by
making uses of the respective spectroscopic parameters of the
models. Our theoretical predictions of the decays of the $c \bar c$,
and $b \bar b$ mesons and the results obtained from some of the
other potential schemes are compared with the experimental values.
The partial widths and life time of the $B_c$ meson are also
computed using the model parameters and are found to be in good
accordance with the experimental values.
\end{abstract}

pacs {12.39Jh, 12.40Yx, 13.20.Gd, 13.20Fc, 13.25.Hw,14.40.Nd}

  \def\be{\begin{equation}}      \def\ol{\overline}
  \def\ee{\end{equation}}    \def\beq{\begin{eqnarray}}
  \def\dis{\displaystyle}    \def\eeq{\end{eqnarray}}
  \def\btd{\bigtriangledown}     \def\m{\multicolumn}
  \def\dfac{\dis\frac}       \def\ra{\rightarrow}

\section{Introduction}
Recently, there have been renewed interest in the spectroscopy of
the heavy flavoured hadrons due to number of experimental facilities
(CLEO, DELPHI, Belle, BaBar, LHCb etc) which have been continuously
providing and expected to provide more accurate and new informations
about the hadrons from low flavour to heavy
flavour sector \cite{ParticleDataGroup2006,particle2002}.\\

The heavy flavour mesons are those in which at least one of the
quark or antiquark or both the quark and antiquark belong to heavy
flavour sector; particularly the charm or beauty. They are
represented by $Q \bar Q$ mesonic systems which include the
quarkonia ($c\bar c$ and $b\bar b$) and $B_c$ ($b\bar c$ or $c\bar
b$) mesons. The investigation of the properties of these mesons
gives very important insight into heavy quark dynamics. Heavy
quarkonia have a rich spectroscopy with many narrow states
lying under the threshold of open flavour production \cite{M.B.Voloshin2008, Eichten2008}.\\

The success  of theoretical model predictions with experiments can
provide important information about the quark-antiquark
interactions. Such information is of great interest, as it is not
possible to obtain the $Q \bar Q$ potential starting from the
basic principle of the quantum chromodynamics (QCD) at the
hadronic scale. In this scale it is necessary to account for
non-perturbative effects connected with complicated structure of
QCD vacuum. All this lead to a theoretical uncertainty in the
$Q\bar Q$ potential at large and intermediate distances. It is
just in this region of large and intermediate distances that most
of the basic hadron resonances are formed. Among many theoretical
attempts or approaches to explain the hadron properties based on
its quark structure very few were successful in predicting the
hadronic properties starting from mass spectra to decay widths.
For the mass predictions, the nonrelativistic potential models
with Buchm\"uller and Tye \cite{BuchmullerTye1981}, Martin
\cite{Martin1980,Amartin1979,Richadson1979}, Log
\cite{Quiggrosner1977,QuiggRosner1979}, Cornell \cite{Eichten1978}
etc., were successful at the heavy flavour sectors while the
Bethe-Salpeter approach under harmonic confinement
\cite{vijayakumar2004} was successful at low flavour sector. There
exist relativistic approaches for the study of the different
hadronic properties \cite{Altarelli1982,Ebert2003}. The
non-relativistic potential model has been successful for $\psi$
and $\Upsilon $ families, while the relativistic approaches yield
better results in the lighter sector. Some potential models have
also predicted the masses and various decays of the heavy-heavy
mesons which are in fair agreement with the experimental results
\cite{SNGupta1996,SGogfrey1986,Khadkikar1983,hawng1996,JNPandya2001,Vinodkumar1999,
AKRai2002,AKRai2005,AKRai2006,EL-hady1999}. A comprehensive review
of developments in heavy quarkonium physics is available in ref \cite{N barmbilla2005}.\\

The new role of the heavy flavour studies as the testing ground
for the non-perturbative aspects of QCD, demands extension of
earlier phenomenological potential model studies on quarkonium
masses to their predictions of decay widths with the non-perturbative
approaches like NRQCD.\\

The decay rates of the heavy-quarkonium states into photons and pairs of
 leptons are among the earliest
applications of perturbative quantum chromodynamics (QCD)
\cite{AppelquistTandPolitzerHD1975,BarbieriRandGatto1976}. In
these analysis, it was assumed that the decay rates of the meson
factored into a short-distance part that is related to the
annihilation rate of the heavy quark and antiquark, and
long-distance factor containing all nonperturbative effects of the
QCD. The short-distance factor calculated in terms of the running
coupling constant $\alpha_s (m_{Q})$ of QCD, evaluated at the
scale of the heavy-quark mass $m_{Q}$, while the long-distance
factor was expressed in terms of the meson's nonrelativistic wave
function, or its derivatives, evaluated at origin. In case of
S-wave decays
\cite{BarbieriR1979,BodwinGTandPetrelliA2002,HagiwaraKandKim1981,MeackenziePandLepage1981}
and in case of P-wave decays into photons
\cite{BarbieriRandCaffo1980-1}, the factorization assumption was
supported by explicit calculations at next-to-leading order in
$\alpha_s$. However, no general argument advanced for its validity
in higher orders of perturbation theory. These divergence cast a
shadow over applications of perturbative QCD to the calculation of
annihilation rates of the heavy
quarkonium states.\\

In this context, an elegant effort was provided by the NRQCD
formalism \cite{BodwinandLepage1995-97}. It consists of a
nonrelativistic Schrodinger field theory for the heavy quark and
antiquark that is coupled to the usual relativistic field theory
for light quarks and gluons. NRQCD not only organize calculation
of all orders in $\alpha_s$, but also elaborate systematically the
relativistic corrections to the conventional formula. Furthermore,
it also provides nonperturbative definitions of the long-distance
factors in terms of matrix elements of NRQCD, making it possible
to evaluate them in the numerical lattice calculations. Analyzing
$S$-wave decays within this frame work, it recover, at leading
order in $v^2$, standard factorization formulae, which contain a
single nonperturbative parameter. At next to leading order in
$v^2$, the decay rates satisfy a more general factorization
formula, which contain two additional independent nonperturbative
matrix elements related to their radial wave functions.\\

Our attempt in this paper would be then to study the heavy-heavy
flavour mesons in the charm and beauty sector in a general frame
work of the potential models. The model parameters used for the
predictions of the masses and their radial wave functions would be
used for the study of their decay properties using NRQCD formalism.\\

For completion,  we present a detail analysis of mass spectra of
$c \bar c$, $c \bar b$ and $b \bar b$ mesons in the potential
scheme of coulomb plus power potential $(CPP_{\nu})$ with the
power index ($\nu$), varying from 0.5 to 2. Spin hyperfine and
spin orbit interactions are introduced to get the S-wave and
P-wave masses of the pseudoscalar and vector mesons. We present
details of the non-relativistic treatment of the heavy quarks
along with the computed results in section-2. The decay constants
$f_{P,V}$ of these mesons incorporating QCD corrections up to
$\textit{O} (\alpha_s)$ are presented in section-3. The weak decay
of the $B_c$ meson and its life time is computed in section-4,
while in section-5 we present the details of the computations of
the di-gamma decays of pseudoscalar states
and the leptonic decay widths of the vector
states of the $c \bar c$ and $b \bar b$ quarkonia in the frame
work of the NRQCD formalism as well as other treatments
incorporating different correction terms to the respective decay
widths. Though the NRQCD formalism takes advantage of the fact
that heavy quark mass is much larger than the other energy scales
such as the binding energy scale, $\Lambda_{QCD}$ and
$|\overrightarrow{p}|$, the energy fluctuations of the heavy
quarks of the order of the light energy scale are implemented in
pNRQCD \cite{APineda1998,NoraBrambilla2003,NoraBrambilla2005}.
A comprehensive comparison of the results are presented in this section. Finally
we draw our conclusions in section-6 of this paper.

\section{Nonrelativistic Treatment for $Q \bar Q$ systems}

Even though there are attempts based on the relativistic theory
like the light front approach for the study of the heavy flavoured
quarks, under non-relativistic approximations, they reproduces the
results of the non-relativistic quark-potential models
\cite{Chienhep-ph/0609036}. In the center of mass frame of the
heavy quark-antiquark system, the momenta of quark and antiquark
are dominated by their rest mass $m_{Q, \bar Q}\gg
\Lambda_{QCD}\sim |\vec{p}\ |$, which constitutes the basis of the
non-relativistic treatment. In NRQCD, the velocity of heavy quark
is chosen as the
expansion parameter \cite{hycheng1997}.\\

Hence, for the study of heavy-heavy bound state systems such as $c
\bar c$, $c \bar b$, $b \bar b$, we consider a nonrelativistic
Hamiltonian given by \cite{AKRai2002,AKRai2005,AKRai2006}
\begin{equation}
\label{eq:nlham} H= M + \frac{p^2}{2m} +  V(r)\end{equation} where
 \begin{equation}
\label{eq:mm1} M=m_Q + m_{\bar Q} , \  \ \  \ and \   \  \  \
m=\frac{m_Q \ m_{\bar Q}}{m_Q + m_{\bar Q}}\end{equation} $m_Q$
and $m_{\bar Q}$ are the mass parameters of quark and antiquark
respectively, p is the relative momentum of each quark and $V(r)$
is the quark antiquark potential. Though linear plus coulomb
potential is a successful well studied non-relativistic model for
heavy flavour sector, their predictions for decay widths are not
satisfactory owing to the improper value of the radial wave
function at the origin compared to other models \cite{AKRai2005}.
Recently, we have considered a general power potential with colour
coulomb term of the form
 \be \label{eq:403} V(r)=\frac{-\alpha_c}{r} + A r^\nu \ee
as the static quark-antiquark interaction potential $(CPP_{\nu})$.
Here, for the study of mesons, $\alpha _c = \dis \frac{4}{3}
\alpha_s $, $\alpha _s$ being the strong running coupling
constant, $A$ is the potential parameter and $\nu $ is a general
power, such that the choice, $\nu =1$ corresponds to the coulomb
plus linear potential. This potential belong to the special
choices of the generality of the potentials, $V(r)=- C r^{\alpha}+
D r^{\beta}+V_0$ \cite{Sameer2004-5,Motyka1998,xtsong1991} with
$V_0=0$ $\alpha=-1$, $\beta=\nu$. Choices of the power index in the
range $0.5\leq\nu\leq2.0$ have been explored for the present
study. The different choices of $\nu$ here, correspond to
different potential forms. Thus, the potential parameter $A$ can
also be different numerically and dimensionally for each choices
of $\nu$. The properties of the light-heavy flavour mesons have
been calculated using the Gaussian trial wave function
\cite{AKRai2002}. Masses and decay constant of the light-heavy
systems are found to be in agreement with the experimental results
for the choice of $\nu\approx0.5$. However, in the case of
heavy-heavy systems the predictions of the masses were
satisfactory but the decay constants and decay rates were not
predicted satisfactorily \cite{AKRai2005}. Hence for the present
study of heavy-heavy flavour mesons, we employ the exponential
trial wave function of the hydrogenic type to generate the
Schr$\ddot{o}$dinger mass spectra. Within the Ritz variational
scheme using the trial radial wave function we obtain the
expectation values of the Hamiltonian as ($\langle H \rangle =
E(\mu ,\nu ) \ $)
 \be \label{eq:404}E(\mu, \nu)=M+\frac{\mu^2}{8
m}+\frac{1}{2}\left(-\mu \alpha_c +A \
\frac{\Gamma(\nu+3)}{\mu^{\nu}} \right)\ee Eqn(\ref{eq:404}) gives
the spin average mass of the ground state. For excited states the
trial wave function is multiplied by an appropriate orthogonal
polynomial function such that the excited trial wave function gets
orthonormalized. So, it is straight forward to assume the trial
wave function for the $(n,l)$ state to be the form given by the
hydrogenic radial wave function,
\begin{equation}  \label{eq:wavfun}
R_{nl}(r) = \left(\frac{\mu^3 (n-l-1)!}{2n(n+l)!}\right)^{1/2} \
(\mu \ r)^l \ e^{- \mu r /2} \ L^{2l+1}_{n-l-1}(\mu
r)\end{equation}
 Here, $\mu$ is the variational parameter and $L^{2l+1}_{n-l-1}(\mu r)$
is Laguerre polynomial. For a chosen value of $\nu$, the
variational parameter, $\mu $ is determined for each state using
the virial theorem
\begin{equation}
\label{eq:virial}
 \left<\frac{P^2}{2 m}\right>=\frac{1}{2}\left<\frac{r
 d V}{dr}\right> \end{equation}

As the interaction potential assumed here does not contain the
spin dependent part, Eqn (\ref{eq:404}) gives the spin average
masses of the system in terms of the power index $\nu$. The spin
average mass for the ground state is computed for the values of
$\nu$ from 0.5 to 2.  We have taken the quark mass parameters $
m_b=4.66 \ GeV$ and $ m_c=1.31 \ GeV$. The potential parameter A
are fixed for each choices of $\nu$ so as to get the experimental
ground state masses of $c \bar c$, $c \bar b$ and $b \bar b$
mesons . The parameters and the fitted values of A for different
systems are listed in Table \ref{tab:Parameters}. The experimental
spin average masses are computed from the experimental masses of
the pseudoscalar and vector mesons using the relation,
\begin{equation}
M_{SA}=M_P+\frac{3}{4}(M_V-M_P) \end{equation}

For the $nJ$ state, we compute the spin-average or the center of
weight mass from the respective experimental values as
 \begin{equation}
M_{CW,nJ}=\frac{\sum_{J} 2(2J+1)\ M_{nJ}}{\sum_{J}
2(2J+1)}\end{equation}

The fitted value of $A$ for each case of the power index $\nu$
along with other model parameters are tabulated in Table
\ref{tab:Parameters} for $c \bar c$, $c \bar b$ and $b \bar b$
systems. The ground state center of weight mass of $3.068\ GeV$,
$6.320\ GeV$ and $9.453\ GeV$ are used to fit the $A$ values for
$c \bar c$, $c \bar b$ and $b \bar b$ systems respectively. The
values of $A(\nu)$ thus obtained for each case of mesonic systems
are then used to predict the higher $S$ and $P$-wave masses (See
Tables \ref{tab:spinaverage}-\ref{tab:pwavmassesofbb}). The fitted
values of $A$ for each $\nu$ in the case of the heavy-heavy
flavour mesons are plotted in Fig \ref{fig:avsnu}. It is
interesting to note that all the three plots intersect each other
at $\nu$ equal to 1.1 at the value of the parameter $A$ around
0.151 GeV$^{\nu+1}$. It can also be seen that the parametric
values of $A$ for $c\bar c$ and $c\bar b$ systems are close to
each other, while in the case of $b\bar b$, they are distinctly
different except at $\nu = 1.1$. It reflects the fact that
potential parameter A becomes independent of the distinct energy
scales of these heavy mesons at around $\nu=1.1$.


In the case of $c\bar c$ and $c\bar b$ systems, the values of  the
parameter $A$ are numerically very close to each other in the
range of potential index 0.9 to 1.3. The predicted masses are also
found to be in good agreement with the existing experimental
states in range of power index 0.9 to 1.3 of the potential.  Fig
\ref{fig:r0vsnu} shows the behaviour of $|R_{1S}(0)|$ with the
potential index $\nu$ for all the three ($c\bar c$, $c\bar b$ and
$b\bar b$) mesons. Like other potential model predictions of the
wave functions (at the origin) of $b \bar c$ system lie in between
those of $c\bar c$ and $b\bar b$ systems. We obtained a model
independent relationship similar to the one given by
\cite{Sterett1997} as
\begin{equation}
\left|\psi_{b\bar c}\right|^2\approx\left|\psi_{c \bar
c}\right|^{2(1-q)}\left|\psi_{b\bar b} \right|^{2q}\end{equation}
with $q=0.3$. This relation provides the 1S wave function at the
origin within $2\%$ variation for the choices of the potential
range $0.5\leq\nu\leq2.$ For $2S$ and $3S$ states we find the
relation hold within $5\%$ for all values for $\nu$ studied here.
It is to be noted here that such a scaling law with smaller
percentage variations exist here even though the potential
contains coulomb part.

\begin{table}
\begin{center}
\caption{Parameter A for mesons in (GeV$^{\nu+1}$)} \vspace{0.1in}
\label{tab:Parameters}
\begin{tabular}{lllll}
\hline $\nu$ &A ($c \bar {c}$) & A ($b \bar{c}$)& A ($b \bar {b}$)\\
\hline
0.5&0.276&0.250&0.195\\
0.7&0.225&0.212&0.179\\
0.9&0.185&0.179&0.165\\
1.0&0.167&0.165&0.158\\
1.1&0.151&0.152&0.151\\
1.3&0.124&0.129&0.138\\
1.5&0.101&0.109&0.126\\
1.7&0.082&0.092&0.114\\
1.9&0.067&0.077&0.103\\
2.0&0.060&0.070&0.098\\
\hline
\end{tabular}
\\ $\alpha_c(c \bar {c})=0.40$, $\alpha_c(b \bar {c})=0.34$,
$\alpha_c(b \bar {b})=0.30$, \\ $m_c=1.31 \ GeV $ and $m_b=4.66 \
GeV$
\end{center}
\end{table}

\begin{table}
\begin{center}
\caption{Wave function at the origin ($|R(0)|$)and spin average
masses of S-wave $c \bar {c}$ meson}\label{tab:spinaverage}
\begin{tabular}{lllllcll}
\hline
 State& $\nu$ &${\bar \mu }$ & $|R(0)|$&\multicolumn{1}{c}{$E(\bar \mu )$}
&Exp.&Theory\\
&&$GeV$&$GeV^{3/2}$&(GeV)&(GeV)&(GeV)\\
\hline
&0.5&1.068&0.781&3.068&&\\
&0.7&1.177&0.903&3.068&&\\
&0.9&1.264&1.005&3.068&&\\
1S&1.0&1.300&1.049&3.068&&\\
&1.1&1.335&1.090&3.068&3.068&3.068$^a$\\
&1.3&1.394&1.164&3.068&&3.068$^b$\\
&1.5&1.445&1.228&3.068&&\\
&1.7&1.489&1.285&3.068&&\\
&1.9&1.528&1.336&3.068&&\\
&2.0&1.546&1.360&3.068&&\\
\hline
&0.5&1.057&0.384&3.368&&\\
&0.7&1.242&0.489&3.454&&\\
&0.9&1.403&0.588&3.534&&\\
2S&1.0&1.473&0.632&3.567&&\\
&1.1&1.540&0.676&3.601&3.663&3.662$^a$\\
&1.3&1.660&0.757&3.661&&3.674$^b$\\
&1.5&1.765&0.829&3.713&&\\
&1.7&1.856&0.894&3.756&&\\
&1.9&1.939&0.955&3.796&&\\
&2.0&1.976&0.982&3.814&&\\
\hline
&0.5&1.097&0.271&3.550&&\\
&0.7&1.333&0.363&3.712&&\\
&0.9&1.545&0.453&3.870&&\\
3S&1.0&1.640&0.495&3.940&&\\
&1.1&1.732&0.537&4.012&4.040&4.064$^a$\\
&1.3&1.901&0.618&4.146&&4.073$^b$\\
&1.5&2.051&0.692&4.266&&\\
&1.7&2.186&0.762&4.373&&\\
&1.9&2.309&0.827&4.473&&\\
&2.0&2.365&0.857&4.518&&\\
\hline
\end{tabular}
\end{center}
Ebert(a) $\ra$ \cite{Ebert2003}, Pandya (b) $\ra$
\cite{JNPandya2001}
\end{table}

\begin{table}
\begin{center}
\caption{Derivative of wave function  at the origin ($|R'(0)|$)
and P-wave masses of $c \bar c$ meson}\label{tab:wavepverage}
\begin{tabular}{llllcll}
\hline State& $\nu$ & ${\bar \mu }$ &
$|R'(0)|$&\multicolumn{1}{c}{$E(\bar \mu )$}
&Exp.&Theory\\
&&$GeV$&$GeV^{5/2}$&(GeV)&(GeV)&(GeV)\\
\hline
&0.5&1.024&0.217&3.313&&\\
&0.7&1.180&0.309&3.373&&\\
&0.9&1.331&0.417&3.426&&\\
1P&1.0&1.392&0.467&3.450&&\\
&1.1&1.450&0.517&3.473&3.525&3.526$^a$\\
&1.3&1.553&0.614&3.513&&3.497$^b$\\
&1.5&1.642&0.705&3.547&&\\
&1.7&1.720&0.792&3.576&&\\
&1.9&1.790&0.875&3.603&&\\
&2.0&1.823&0.916&3.615&&\\
\hline
&0.5&1.081&0.405&3.519&&\\
&0.7&1.307&0.651&3.666&&\\
&0.9&1.509&0.932&3.808&&\\
2P&1.0&1.560&1.013&3.872&&\\
&1.1&1.687&1.232&3.936&&3.945$^a$\\
&1.3&1.847&1.545&4.055&&3.907$^b$\\
&1.5&1.990&1.862&4.162&&\\
&1.7&2.117&2.174&4.257&&\\
&1.9&2.232&2.481&4.345&&\\
&2.0&2.285&2.631&4.385&&\\
\hline
\end{tabular}
\end{center}
Ebert(a) $\ra$ \cite{Ebert2003}, Pandya (b) $\ra$
\cite{JNPandya2001}
\end{table}

\begin{table}
\begin{center}\caption{Wave function  at the origin ($|R(0)|$)and spin average masses of S-wave
$b \bar {c}$ meson}\label{tab:vefunction}
\begin{tabular}{cccccc}
\hline State& $\nu$ & ${ \bar \mu }$ &
$|R(0)|$&\multicolumn{1}{c}{$E(\bar \mu ) $}
&Theory\\
&&$GeV$&$GeV^{3/2}$&(GeV)&(GeV)\\
\hline
&0.5&1.281&1.025&6.320&\\
&0.7&1.404&1.176&6.320&\\
&0.9&1.502&1.301&6.320&\\
1S&1.0&1.544&1.357&6.320&6.317$^a$\\
&1.1&1.583&1.408&6.320&6.319$^c$\\
&1.3&1.652&1.502&6.320&\\
&1.5&1.711&1.583&6.320&\\
&1.7&1.763&1.655&6.320&\\
&1.9&1.807&1.718&6.320&\\
&2.0&1.825&1.744&6.320&\\
\hline
&0.5&1.236&0.486&6.589&\\
&0.7&1.453&0.619&6.665&\\
&0.9&1.635&0.739&6.730&\\
2S&1.0&1.719&0.797&6.761&\\
&1.1&1.796&0.851&6.790&\\
&1.3&1.938&0.954&6.844&6.869$^a$\\
&1.5&2.061&1.046&6.890&6.888$^c$\\
&1.7&2.172&1.132&6.931&\\
&1.9&2.266&1.206&6.965&\\
&2.0&2.307&1.239&6.977&\\
\hline
&0.5&1.274&0.339&6.746&\\
&0.7&1.550&0.455&6.889&\\
&0.9&1.792&0.565&7.021&\\
3S&1.0&1.905&0.620&7.085&\\
&1.1&2.012&0.674&7.147&\\
&1.3&2.211&0.775&7.265&7.224$^a$\\
&1.5&2.389&0.870&7.372&7.271$^c$\\
&1.7&2.550&0.960&7.471&\\
&1.9&2.692&1.041&7.555&\\
&2.0&2.755&1.078&7.590&\\
\hline
\end{tabular}
\end{center}
$^a$Ebert $\ra$ \cite{Ebert2003}; $^c$Eichten $\ra$
\cite{Eichten1994}
\end{table}

\begin{table}
\begin{center}\caption{Derivative of wave function at the origin
($|R'(0)|$) and P-wave masses of $c \bar b$
meson}\label{tab:avefunion}
\begin{tabular}{cccccc}
\hline\hline State& $\nu$& ${ \bar \mu }$ &
$|R'(0)|$&\multicolumn{1}{c}{$E(\bar \mu ) $}
&Theory\\
&&$GeV$&$GeV^{5/2}$&(GeV)&(GeV)\\
\hline
&0.5&1.198&0.321&6.542&\\
&0.7&1.392&0.467&6.597&\\
&0.9&1.553&0.613&6.641&\\
1P&1.0&1.625&0.687&6.662&\\
&1.1&1.692&0.760&6.682&\\
&1.3&1.814&0.905&6.718&6.749$^a$\\
&1.5&2.920&1.042&6.749&6.736$^c$\\
&1.7&2.013&1.173&6.777&\\
&1.9&2.094&1.296&6.799&\\
&2.0&2.129&1.349&6.806&\\
\hline
&0.5&1.260&0.594&6.720&\\
&0.7&1.520&0.950&6.851&\\
&0.9&1.751&1.353&6.969&\\
2P&1.0&1.859&1.571&7.027&\\
&1.1&1.961&1.794&7.082&\\
&1.3&2.149&2.257&7.188&7.145$^a$\\
&1.5&2.317&2.725&7.283&7.142$^c$\\
&1.7&2.469&3.184&7.370&\\
&1.9&2.603&3.644&7.445&\\
&2.0&2.662&3.852&7.476&\\
\hline
\end{tabular}
\end{center}
$^a$Ebert $\ra$ \cite{Ebert2003}; $^c$Eichten $\ra$
\cite{Eichten1994}
\end{table}

\begin{table}
\begin{center}\caption{Wave function at the origin  ($|R(0)|$)and spin average masses of S-wave
$b \bar {b}$ meson}\label{tab:sbbmasses}
\begin{tabular}{cccccc}
\hline State&$\nu$&${\bar \mu }$&$|R(0)|$&\multicolumn{1}{c}{$E(\bar \mu)$}&Theory\\
&&$GeV$&$GeV^{3/2}$&(GeV)&(GeV)\\
\hline
&0.5&1.985&1.977&9.453&\\
&0.7&2.122&2.186&9.453&\\
&0.9&2.238&2.368&9.453&\\
1S&1.0&2.288&2.447&9.453&\\
&1.1&2.336&2.525&9.453&9.445$^a$\\
&1.3&2.402&2.662&9.453&9.453$^d$\\
&1.5&2.491&2.780&9.453&\\
&1.7&2.554&2.885&9.453&\\
&1.9&2.611&2.982&9.453&\\
&2.0&2.638&3.030&9.453&\\
\hline
&0.5&1.701&0.784&9.701&\\
&0.7&1.979&0.984&9.758&\\
&0.9&2.227&1.175&9.812&\\
2S&1.0&2.338&1.264&9.838&\\
&1.1&2.442&1.349&9.861&10.016$^a$\\
&1.3&2.636&1.513&9.905&10.008$^d$\\
&1.5&2.807&1.663&9.944&\\
&1.7&2.958&1.790&9.977&\\
&1.9&3.094&1.924&10.008&\\
&2.0&3.158&1.984&10.023&\\
\hline
&0.5&1.692&0.519&9.826&\\
&0.7&2.053&0.694&9.935&\\
&0.9&2.385&0.868&10.043&\\
3S&1.0&2.538&0.953&10.095&\\
&1.1&2.684&1.036&10.144&10.348$^a$\\
&1.3&2.957&1.199&10.241&10.351$^d$\\
&1.5&3.205&1.353&10.330&\\
&1.7&3.427&1.496&10.410&\\
&1.9&3.631&1.631&10.485&\\
&2.0&3.727&1.696&10.522&\\
\hline
\end{tabular}
\end{center}
 Ebert (a) $\ra$ \cite{Ebert2003}, Gupta (d) $\ra$ \cite{SNGuPta1982}
\end{table}

\begin{table}
\begin{center}\caption{Derivative of wave function at the origin  ($|R'(0)|$) and P-wave masses of $b \bar b$
meson}\label{tab:pwavmassesofbb}
\begin{tabular}{ccccccc}
\hline\hline State& $\nu$ & ${\bar \mu }$ &
$|R'(0)|$&\multicolumn{1}{c}{$E(\bar \mu ) $}
&Exp&Theory\\
&&$GeV$&$GeV^{5/2}$&(GeV)&(GeV)&(GeV)\\
\hline
&0.5&1.654&0.718&9.670&&\\
&0.7&1.904&1.021&9.712&&\\
&0.9&2.121&1.337&9.751&&\\
1P&1.0&2.218&1.446&9.768&&\\
&1.1&2.308&1.653&9.784&9.900&9.901$^a$\\
&1.3&2.474&1.966&9.815&&9.900$^d$\\
&1.5&2.621&2.271&9.842&&\\
&1.7&2.750&2.559&9.864&&\\
&1.9&2.866&2.838&9.885&&\\
&2.0&2.921&2.977&9.896&&\\
\hline
&0.5&1.669&1.200&9.808&&\\
&0.7&2.016&1.922&9.908&&\\
&0.9&2.333&2.770&10.006&&\\
2P&1.0&2.478&3.223&10.053&&\\
&1.1&2.617&3.692&10.097&10.260&10.261$^a$\\
&1.3&2.876&4.677&10.184&&10.258$^d$\\
&1.5&3.111&5.691&10.264&&\\
&1.7&3.321&6.699&10.335&&\\
&1.9&3.513&7.708&10.401&&\\
&2.0&3.604&8.218&10.434&&\\
\hline
\end{tabular}
\end{center}
Ebert (a) $\ra$ \cite{Ebert2003}, Gupta (d) $\ra$ \cite{SNGuPta1982}
\end{table}

\subsection{Spin-Hyperfine and Spin-Orbit Splitting in Heavy-Heavy Flavour Mesons}
In general, the quark-antiquark bound states are represented by
$n^{2S+1} L_J$,  identified with the $J^{PC}$ values, with $ \vec
J=\vec L + \vec S$, $ \vec S=\vec
 S_Q + \vec S_{\bar Q}$, parity $P=(-1)^{L+1}$ and the charge
conjugation $C = (-1)^{L+S}$ and (n, L) is the radial quantum
numbers. So the $S$-wave $(L=0)$ bound states are represented by
$J^{PC}=0^{-+}$ and $1^{--}$ respectively. And the $P$-wave $(L=1)$
states are represented by $J^{PC}=1^{+-}$ with $L=1$ and $S=0$ while
$J^{PC}=0^{++},\ 1^{++}$ and $2^{++}$ correspond to $L=1$ and $S=1$
respectively. Accordingly, the spin-spin interaction among the
constituent quarks provides the mass splitting of $J=0^{-+}$ and
$1^{--}$ states, while the spin-orbit interaction provides the mass
splitting of $J^{PC}=0^{++},\ 1^{++}$ and $2^{++}$ states. The
$J^{PC} = 1^{+-}$ state with $L=1$ and $S=0$ represents the center
of weight mass of the $P$-state as its spin-orbit contribution
becomes zero, while the two $J=1^{+-}$ singlet and  the $J=1^{++}$
of the triplet P-states form a mixed sate. We add separately the
spin-dependent part of the usual one gluon exchange potential (OGEP)
between the quark antiquark for computing the hyperfine and
spin-orbit shifting of the low-lying $S$-states and $P$-states.
Accordingly, the spin-spin and spin-orbit interactions are taken as
\cite{Gerstein1995}
\begin{equation} \label{spindependent}
 V_{S_Q \ \cdot \ S_{\bar Q}}(r)= \frac{8}{9} \frac{ \alpha_s}{ m_Q m_{\bar Q}}
 \ \vec S_Q \ \cdot  \vec S_{\bar Q} \ 4 \pi \delta(r)\end{equation}
\begin{equation} \label{spinorbit}
V_{L\ \cdot \ S}(r)= \frac{4 \ \alpha_s}{3 \ m_Q m_{\bar Q}}
\frac{\vec L \ \cdot  \vec S}{r^3}\end{equation} The value of the
radial wave function R(0) for $0^{- \ +}$ and $1^{--}$ states
would be different due to their spin dependent hyperfine
interaction. The spin hyperfine interaction of the heavy flavour
mesons are small and this can cause a small shift in the value of
the wave function at the origin. Thus, many other models do not
consider this contribution to their value of R(0). However, we
account this correction to the value of R(0) by considering
\be \label{wavefunctionhy}
R_{nJ}(0)=R(0)\left[1+(SF)_J
\frac{<\varepsilon_{SD}>_{nJ}}{M_{SA}} \right]\ee Where $(SF)_J$
and $<\varepsilon_{SD}>_{nJ}$ is the spin factor and spin
interaction energy of the meson in the $nJ$ state, while $R(0)$
and $M_{SA}$ correspond to the radial wave function at
the zero separation and spin average mass respectively of the $Q \bar Q$ system. It can be seen that Eqn(\ref{wavefunctionhy}) provides the average radial wave function given by \cite{BodwinandLepage1995-97} as
\be
R(0)=\frac{R_p + 3 R_v}{4}\ee

It is found that the computed mass increases with increase of $\nu$.
The computed results for the pseudoscalar($P$) and Vector ($V$)
mesons in the case of $c \bar c, \ c \bar b, \ b \bar b$ systems are
tabulated in Tables \ref{tab:ccgrdst}-\ref{tab:bbspect}. The
spin-spin hyperfine and spin-orbit interactions are computed
perturbatively to get the masses of $\eta_c$, $J/\psi$, $B_c$,
$B_c^*$, $\eta_b$, and $\Upsilon$ states. The results are compared
with known experimental values as well as with other theoretical
predictions. Mass predictions with $\nu$ between $1.0$ and $1.5$
are found in accordance with the experimental results \cite{particle2002}.\\
{\tiny
\begin{table*}
\begin{center}
\caption{S-Wave and P-Wave Masses (in $GeV$) of $c \bar c$ meson}
\label{tab:ccgrdst}
\begin{tabular}{ccccccccccccccc}
\hline $\nu$&$1^1S_0$&$1\ ^3S_1$&$1^1P_1$&1\ $^3 P_0$&1\ $^3 P_1$&1\
$^3 P_2$&$2^1S_0$&2\
$^3S_1$&$2^1P_1$&2$^3 P_0$&2\ $^3 P_1$&2\ $^3P_2$&$3^1S_0$&3\ $^3S_1$ \\
\hline
0.5&3.000&3.092&3.313 &3.292 &3.302&3.323&3.352&3.375&3.519&3.494&3.507&3.531&3.541&3.553\\
0.7&2.980&3.100&3.373 &3.341 &3.357&3.389&3.427&3.464&3.666&3.623&3.644&3.687&3.697&3.717\\
0.9&2.960&3.109&3.429 &3.383 &3.406&3.451&3.495&3.547&3.808&3.742&3.775&3.842&3.846&3.878\\
1.0&2.950&3.112&3.450 &3.398 &3.424&3.477&3.522&3.583&3.872&3.792&3.832&3.911&3.912&3.950\\
1.1&2.942&3.116&3.473 &3.414 &3.444&3.503&3.549&3.619&3.936&3.843&3.889&3.983&3.979&4.024\\
1.3&2.926&3.123&3.513 &3.441 &3.477&3.550&3.597&3.683&4.055&3.933&3.994&4.116&4.102&4.161\\
1.5&2.912&3.129&3.547 &3.461 &3.504&3.590&3.636&3.739&4.162&4.009&4.085&4.239&4.212&4.285\\
1.7&2.899&3.134&3.576 &3.477 &3.526&3.625&3.668&3.788&4.257&4.073&4.165&4.394&4.309&4.395\\
1.9&2.887&3.141&3.603 &3.491 &3.547&3.658&3.696&3.832&4.345&4.129&4.237&4.453&4.396&4.500\\
2.0&2.882&3.144&3.615 &3.497 &3.556&3.673&3.708&3.852&4.385&4.153&4.269&4.501&4.436&4.547\\
\hline
\cite{particle2002,ParticleDataGroup2006}&2.980&3.097&3.511&3.415&&3.556&3.622\cite{Abe2002}&3.686&&&&3.929&&4.040\\
\hline
\cite{Ebert2003}&2.979&3.096&3.526&3.424&3.511&3.556&3.588&3.686&3.945&3.854&3.929&3.972&3.991&4.088\\
\hline
\cite{sfredford2007}&2.980&3.097&3.527&3.416&3.508&3.558&3.597&3.686&3.960&3.844&3.894&3.994&4.014&4.095\\
\hline
\end{tabular}\\
\end{center}
\end{table*}}

\begin{table*}
\begin{center}
\caption{S-Wave and P-Wave Masses (in $GeV$) of $c\bar b$ meson}
\label{tab:bcspectrm}
\begin{tabular}{ccccccccccccccc}
\hline $\nu$&$1^1S_0$&$1\ ^3S_1$&$1^1P_1$&1\ $^3 P_0$&1\ $^3 P_1$&1\
$^3 P_2$&$2^1S_0$&2\
$^3S_1$&$2^1P_1$&2$^3 P_0$&2\ $^3 P_1$&2\ $^3P_2$&$3^1S_0$&3\ $^3S_1$ \\
\hline
0.5&6.291&6.330&6.542&6.534&6.538&6.546&6.582&6.591&6.720&6.711&6.715&6.726&6.743&6.747\\
0.7&6.283&6.334&6.597&6.584&6.590&6.603&6.655&6.669&6.851&6.834&6.840&6.859&6.884&6.891\\
0.9&6.273&6.335&6.641&6.624&6.633&6.650&6.715&6.735&6.969&6.944&6.957&6.982&7.012&7.024\\
1.0&6.269&6.337&6.662&6.642&6.652&6.672&6.743&6.767&7.027&6.997&7.012&7.042&7.075&7.089\\
1.1&6.265&6.338&6.682&6.659&6.671&6.693&6.770&6.797&7.082&7.047&7.065&7.099&7.135&7.151\\
1.3&6.259&6.341&6.718&6.691&6.704&6.732&6.819&6.852&7.188&7.142&7.165&7.211&7.249&7.271\\
1.5&6.252&6.344&6.749&7.716&7.733&7.765&6.860&6.900&7.283&7.225&7.254&7.312&7.351&7.379\\
1.7&6.247&6.347&6.777&7.739&7.758&7.796&6.896&6.943&7.370&7.300&7.335&7.405&7.445&7.479\\
1.9&6.241&6.348&6.799&7.756&7.777&7.820&6.924&6.978&7.445&7.363&7.404&7.486&7.525&7.565\\
2.0&6.237&6.348&6.806&7.762&7.784&7.829&6.935&6.991&7.476&7.388&7.432&7.519&7.558&7.601\\
\hline
\cite{Ebert2003}&6.270&6.332&6.749&6.699&6.734&6.762&6.835&6.881&7.145&7.091&7.126&7.145&7.193&7.235\\
\hline
\cite{Eichten1994}&6.256&6.337&6.755&6.700&6.730&6.747&6.899&6.929&7.169&7.108&7.135&7.142&7.280&7.308\\
\hline
\end{tabular}\\
\end{center}
\end{table*}

\begin{table*}
\begin{center}
\caption{S-Wave and P-Wave Masses (in $GeV$) of $b \bar b$ meson}
\label{tab:bbspect}
\begin{tabular}{ccccccccccccccc}
\hline $\nu$&$1^1S_0$&$1\ ^3S_1$&$1^1P_1$&1\ $^3 P_0$&1\ $^3 P_1$&1\
$^3 P_2$&$2^1S_0$&2\
$^3S_1$&$2^1P_1$&2$^3 P_0$&2\ $^3 P_1$&2\ $^3P_2$&$3^1S_0$&3\ $^3S_1$ \\
\hline
0.5&9.426&9.463&9.672&9.664&9.670&9.683&9.696&9.702&9.808&9.803&9.806&9.811&9.824&9.827\\
0.7&9.419&9.465&9.716&9.703&9.712&9.731&9.751&9.760&9.908&9.898&9.903&9.913&9.931&9.936\\
0.9&9.414&9.467&9.757&9.740&9.751&9.774&9.803&9.816&10.006&9.991&9.999&10.014&10.038&10.045\\
1.0&9.411&9.468&9.775&9.755&9.768&9.792&9.826&9.841&10.053&10.035&10.044&10.062&10.088&10.097\\
1.1&9.408&9.468&9.791&9.769&9.784&9.809&9.846&9.865&10.097&10.076&10.086&10.108&10.136&10.147\\
1.3&9.403&9.470&9.824&9.797&9.815&9.840&9.888&9.910&10.184&10.155&10.170&10.198&10.230&10.244\\
1.5&9.399&9.472&9.852&9.820&9.842&9.866&9.924&9.951&10.264&10.228&10.246&10.282&10.317&10.334\\
1.7&9.394&9.473&9.877&9.840&9.864&9.887&9.955&9.985&10.335&10.291&10.313&10.357&10.394&10.416\\
1.9&9.390&9.474&9.900&9.857&9.885&9.905&9.982&10.017&10.401&10.350&10.376&10.428&10.466&10.492\\
2.0&9.389&9.475&9.911&9.866&9.896&9.913&9.995&10.032&10.434&10.379&10.406&10.462&10.501&10.529\\
\hline \hline
\cite{particle2002,ParticleDataGroup2006}&&9.460&&9.860&9.893&9.913&&10.023&&10.232&10.255&10.268&&10.355\\
\hline
\cite{Ebert2003}&9.400&9.460&9.901&9.863&9.892&9.913&9.993&10.023&10.261&10.234&10.255&10.268&10.328&10.355\\
\hline
\cite{sfredford2007}&9.414&9.461&9.900&9.861&9.891&9.912&9.999&10.023&10.262&10.231&10.255&10.272&10.345&10.364\\
\hline
\end{tabular}\\
\end{center}
\end{table*}

\section{Decay constants ($f_{P/V}$) of the heavy flavoured mesons}
The decay constants of mesons are important parameters in the
study of leptonic or non-leptonic weak decay processes. The decay
constants  of pseudoscalar ($f_P$) and vector ($f_V$) mesons are
obtained by prarameterizing  the matrix elements of weak current
between the corresponding mesons and the vacuum as \be \langle 0 |
\bar Q \gamma^\mu \gamma_5 Q | P_{\mu}(k)\rangle = i f_P k^\mu \ee
\be \langle 0 | \bar Q \gamma^\mu Q | V(k,\epsilon)\rangle = f_V
M_V \epsilon^\mu \ee

where $k$ is the meson momentum, $\epsilon^\mu$ and $M_V$ are the
polarization vector and mass of the vector meson. In the
relativistic quark model, the decay constant can be expressed
through the meson wave function $\Phi_{P,V}(p)$ in the momentum
space as \cite{Ebert2003}

 \beq f_{P,V}  =
\dis\sqrt{\frac{12}{M_{P,V}}}\int\dis\frac{d^3p}{(2\pi)^3}\sqrt{
\left ( \frac{E_Q(p) + m_Q}{2E_Q(p)}\right )} \sqrt{\left (
\frac{E_{\bar Q}(p) + m_{\bar Q}}{2E_{\bar Q}(p)}\right )} \left
\{1+ \lambda_{P,V} \frac{p^2}{[E_Q(p)+m_Q][E_{\bar Q}(p)+m_{\bar
Q}]}\right \}\Phi_{P,V}(p) \label{tab:pseudj}\eeq

with $\lambda_P=-1$ and $\lambda_V=-1/3$. In the nonrelativistic
limit $\dis\frac{p^2}{m^2} \ra 0$, this expression reduces to the
well known relation between  $f_{P,V}$ and the ground state wave
function at the origin $\psi_{P,V}(0)$, the Van-Royen-Weisskopf
formula \cite{Vanroyenaweissskopf}. Though most of the models
predict the mesonic mass spectrum successfully, there are
disagreements in the predictions of the pseudoscalar and vector
decay constants. For example, most of the cases, the ratio
$\frac{f_P}{f_V}$ was predicted to be $>1$ as $m_P<m_V$ and their
wave function at the origin $\psi_P(0) \sim \psi_V(0)$ \cite{1997}.
The ratio computed in the relativistic models \cite{1997} predicted
the ratio $\frac{f_P}{f_V}<1$, particularly in the heavy flavours
sector. The disparity of the predictions of these decay constants
play decisive role in the precision measurements of the weak decay
parameters. So, we re-examine the predictions of the decay constants
under different potential schemes discussed in the present work.
Incorporating a first order QCD correction factor , we compute them
using the relation,

\be f^2_{P/V}=\frac{12 \left| \psi_{P/V}(0)\right|^2} {M_{P/V}} {\bar
C}(\alpha_s)\label{eq:414} \ee Where ${\bar C}(\alpha_s)$ is the QCD correction
factor given by \cite{EBraaten1995,Gershtein1998} \be {\bar
C}(\alpha_s)=1-\frac{\alpha_s}{\pi} \left[\delta^{P,V}-\frac{m_Q-m_{\bar
Q}}{m_Q+m_{\bar Q}} \ ln \frac{m_Q}{m_{\bar Q}} \right] \label{eq:fpv}\ee

Here $\delta^{P}=2$ and $\delta^{V}=8/3$. The computed $f_{P}$ and $f_{V}$ for $c \bar
c$, $c \bar b$ and $b \bar b $ systems using Eqn(\ref{eq:414}) \& (\ref{eq:fpv}) and
the
 predicted radial wave functions at the origin $R_{nJ}(0)$ of the respective
mesons are tabulated in Tables \ref{tab:etacandjpsi} -
\ref{tab:etaandupsilon}. The decay constants without and with the
QCD corrections are also listed as $f_{P,V}$ and $f_{P,V}(cor.)$
in the table. The plot of $f_P\ vs \ (m_Q+m_{\bar Q})$ shows (see
fig 3) deviations from linearity as against the predictions of a
linear scaling between the weak decay constant and the sum of
quark antiquark masses justified within a renormalized light front
QCD inspired model for quark antiquark bound states
\cite{Saldo2003}.

\begin{table}
\begin{center}
\caption{Decay constants($f_P$ $\&$ $f_V$) (in MeV) of $1S$ and
 $c\bar{c}$ mesons states.}\label{tab:etacandjpsi}
\begin{tabular}{ccccccc}
\hline Models&$R_p(0)$&$R_v(0)$&$f_P$
&$f_P(cor.)$&$f_{V}$&$f_{V}(cor.)$ \\
&$GeV^{3/2}$&$GeV^{3/2}$&$MeV$&$MeV$&$MeV$&$MeV$\\
\hline
ERHM&0.726&0.752&410&317&418&323\\
BT&0.874&0.909&499&382&505&389\\
PL&0.971&1.009&550&399&560&407\\
LOG&0.877&0.914&496&379&506&387\\
Cornell&1.171&1.217&663&532&676&543\\
&&&&&&\\
$\nu=$0.5   &   0.763   &   0.787   &   430 &   348 &   437 &   326 \\
0.7 &   0.875   &   0.912   &   495 &   401 &   506 &   377 \\
0.9 &   0.967   &   1.018   &   549 &   444 &   564 &   421 \\
1.0 &   1.005   &   1.063   &   572 &   463 &   589 &   439 \\
1.1 &   1.041   &   1.107   &   593 &   480 &   613 &   457 \\
1.3 &   1.104   &   1.184   &   627 &   507 &   655 &   488 \\
1.5 &   1.158   &   1.252   &   663 &   536 &   692 &   516 \\
1.7 &   1.204   &   1.311   &   691 &   559 &   724 &   539 \\
1.9 &   1.245   &   1.366   &   716 &   579 &   753 &   561 \\
2.0 &   1.264   &   1.391   &   728 &   589 &   767 &   571 \\

\hline
&&& &335$\pm$&459$\pm$&416$\pm$\\
&&&& 75\cite{Cleocollaboration2001}& 28\cite{glwang206}&6\cite{Cleocollaboration2001}\\
\hline
\end{tabular}
\end{center}
\end{table}

\begin{table}
\begin{center}
\caption{Pseudoscalar and Vector meson decay constants($f_P$ $\&$
$f_V$) (in MeV) of $1S$ $b\bar{c}$ meson
state.}\label{tab:pdcmcbc}
\begin{tabular}{cccccccc}
\hline &$R_p(0)$&$R_v(0)$&$f_P$
&$f_P(cor.)$&$f_{V}$&$f_{V}(cor.)$ \\
&&&&&&\\
\hline
$\nu=$0.5   &   1.021   &   1.027   &   398 &   356 &   399 &   336 \\
0.7 &   1.169   &   1.178   &   456 &   408 &   457 &   385 \\
0.9 &   1.291   &   1.304   &   504 &   451 &   506 &   426 \\
1.0 &   1.346   &   1.361   &   525 &   470 &   528 &   445 \\
1.1 &   1.396   &   1.413   &   545 &   488 &   548 &   461 \\
1.3 &   1.487   &   1.507   &   581 &   520 &   585 &   492 \\
1.5 &   1.565   &   1.589   &   612 &   548 &   616 &   519 \\
1.7 &   1.635   &   1.662   &   639 &   572 &   645 &   542 \\
1.9 &   1.695   &   1.725   &   663 &   594 &   669 &   563 \\
2.0 &   1.722   &   1.758   &   674 &   603 &   682 &   574 \\

\hline
&&&&433\cite{Ebert2003}&503\cite{Ebert2003}&418\\
&&&&&& $\pm$ 24\cite{glwang206}\\
\hline
\end{tabular}
\end{center}
\end{table}

\begin{table}
\begin{center}
\caption{Decay constants($f_P$ $\&$ $f_V$) (in MeV) of $1S$ $b
\bar{b}$ meson state.}\label{tab:etaandupsilon}
\begin{tabular}{ccccccc}
\hline Models&$R_p(0)$&$R_v(0)$&$f_P$
&$f_P(cor.)$&$f_{V}$&$f_{V}(cor.)$ \\
\hline
&$GeV^{3/2}$&$GeV^{3/2}$&$MeV$&$MeV$&$MeV$&$MeV$\\
\hline
ERHM&2.232&2.235&709&601&710&601\\
BT&2.527&2.551&807&683&810&686\\
PL&2.132&2.146&680&563&682&565\\
LOG&2.206&2.221&703&594&706&596\\
Cornell&3.706&3.762&1185&1022&1194&1029\\
&&&&&&\\
$\nu=$0.5   &   1.971   &   1.979   &   627 &   537 &   629 &   509 \\
0.7 &   2.178   &   2.189   &   693 &   594 &   695 &   563 \\
0.9 &   2.358   &   2.371   &   751 &   643 &   753 &   609 \\
1.0 &   2.436   &   2.451   &   776 &   665 &   778 &   630 \\
1.1 &   2.513   &   2.529   &   801 &   686 &   803 &   650 \\
1.3 &   2.648   &   2.667   &   844 &   723 &   847 &   685 \\
1.5 &   2.764   &   2.785   &   881 &   755 &   884 &   715 \\
1.7 &   2.867   &   2.891   &   914 &   783 &   918 &   743 \\
1.9 &   2.962   &   2.989   &   945 &   809 &   949 &   768 \\
2.0 &   3.009   &   3.037   &   960 &   822 &   964 &   780 \\

\hline
&&&&711\cite{JNPandya2001}&&708\\
&&&&&&$\pm$ 8\cite{Chienhep-ph/0609036}\\
\hline
\end{tabular}\\
\end{center}
\end{table}



\section{Weak decay of $B^+_c$ meson}
The Decay properties of $B^+_c$ ($\bar b c$) meson is of interest
as it decays only through weak interactions
\cite{Ebert2003,EL-hady1999,sg,CDFcollabration}. This is due to
the fact that its ground state energy lies below the (BD)
threshold and has non vanishing flavour. This eliminates the
uncertainties encountered due to strong decays and provides a
clear decay width and lifetime for $B^+_c$ meson, which helps to
fix more precise value of the weak decay parameters such as the
CKM mixing matrix elements ($V_{cb},\ V_{cs}$) and the leptonic
decay constant ($f_p$). Adopting the spectator model for the
charm-beauty system \cite{EL-hady1999}, the total decay width of
$B^+_c$ meson can be approximated as the sum of the widths of
$\bar b$-quark decay keeping $c$-quark as spectator, the $c$-quark
decay with $\bar b$-quark as spectator, and the annihilation
channel $B^+_{c} \rightarrow l^+\nu_l (c \bar s, \ u \bar s)$,
$l=e,\mu,\tau$ with no interference assumed between them.
\begin{table}
\begin{center}\caption{Decay widths (in $10^{-4}\ eV$) and lifetime $\tau$ (in ps) of $B^+_{c}$
meson} \label{ta:weakdcy} \vspace{0.01cm}
\begin{tabular}{ccccccc}
\m{7}{c}{}\\
\hline Model&\m{2}{c}{$\Gamma(Anni)$}&\m{2}{c}{$\Gamma(B_c
\rightarrow X)$}&\m{2}{c}{$\tau$ (in PS)}\\
\hline
&a&b&a&b&a&b\\
\hline
$\nu=0.5$&0.370&0.370&12.47&13.18&0.530&0.499\\
\ \ \ \ \ 0.7&0.486&0.484&12.59&13.30&0.523&0.495\\
\ \ \ \ \ 0.9&0.596&0.593&12.70&13.41&0.518&0.491\\
\ \ \ \ \ 1.0&0.644&0.642&12.75&13.46&0.516&0.489\\
\ \ \ \ \ 1.1&0.693&0.690&12.79&13.51&0.515&0.487\\
\ \ \ \ \ 1.3&0.786&0.783&12.89&13.60&0.511&0.484\\
\ \ \ \ \ 1.5&0.871&0.867&12.89&13.68&0.507&0.481\\
\ \ \ \ \ 1.7&0.951&0.950&12.97&13.76&0.504&0.478\\
\ \ \ \ \ 1.9&1.023&1.020&13.05&13.83&0.502&0.476\\
\ \ \ \ \ 2.0&1.053&1.050&13.15&13.87&0.500&0.475\\
\hline
\cite{ParticleDataGroup2006}&&&&&$0.46^{+0.18}_{-0.16}$&\\
\hline
\cite{EL-hady1999}&&1.40&&14.00&0.47&\\
\hline
\cite{sg}&& 0.67&&8.8&0.75&\\
\hline
\end{tabular}\\
\end{center}
\end{table}

Accordingly, the total width is written as \cite{EL-hady1999}
 \be
\label{eq:totalwidth} \Gamma(B_c \rightarrow X)= \Gamma(b
\rightarrow X)+ \Gamma(c \rightarrow X)
 + \Gamma(Anni)\ee
Neglecting the quark binding effects, we obtain for the b and c
inclusive widths in the spectator approximation as
\cite{EL-hady1999}
 \beq
 \Gamma(b \rightarrow X)&=& \frac{ 9 \ G^2_F \left|V_{cb}\right|^2
 m^5_b}{192 \pi^3}\cr && =7.97\times 10^{-4} eV (a) \cr && =8.66\times 10^{-4} eV (b)
 \eeq
 \beq\label{restol}
 \Gamma(c \rightarrow X) &=& \frac{ 5 \ G^2_F \left|V_{cs}\right|^2
 m^5_c}{192 \pi^3} \cr && =4.13\times 10^{-4} eV (a) \cr && =4.15\times 10^{-4} eV (b)
 \eeq
Here we have used the model quark masses and the two values (a) and (b) correspond to
the two set of values for the CKM matrix elements
(a)$\rightarrow$$\left|V_{cs}\right|=0.97296$, $\left|V_{cb}\right|=0.04221$ as used in
reference \cite{ParticleDataGroup2006} and (b)$\rightarrow$
$\left|V_{cs}\right|=0.975$, $\left|V_{cb}\right|=0.044$ as the upper bound provided by
particle data group. The values of $ \Gamma(B \rightarrow X)$ and $\Gamma(c \rightarrow
X)$ in Bethe-Salpeter model \cite{EL-hady1999} and relativized quark model \cite{sg}
are 7.5 \& 5.1 and 4.8 \& 3.3 (widths are in $10^{-4}$ eV) respectively. \\

Employing the computed mass of the $1 ^1S_0$ state ($M_{B_c}$) and
$f_{B_c}$ values obtained from the present study, the width of the
annihilation channel is computed using the expression given by
\cite{EL-hady1999}, \be\label{ttlannrest}
  \Gamma(Anni)= \frac{G^2_F}{8 \pi} \left|V_{bc}\right|^2 f^2_{B_C} M_{bc}
 m^2_i\left(1- \frac{m^2_q}{M^2_{B_c}}\right)^2 \  \ C_q, \ee
Where $C_q=3\left|V_{cs}\right|^2$ for $c \bar s$, and $m_q$ is
the mass of the heaviest fermions.\\

The computed widths and lifetime in our $CPP_{\nu}$ model are listed in Table
\ref{ta:weakdcy}. Our predictions for the life time with the potential index
$0.5\succeq \nu \succeq 2$ lie well within the experimental error bar.

\section{Decay rates of quarkonia}
Along with the mass spectrum, successful predictions of various
decay widths of heavy flavoured systems have remained as testing
ground for the success of phenomenological models. Experimentally,
the excited states and the leptonic, di-gamma and other hadronic
decay width, of the heavy flavour mesons have been reported.
However, experimentally, the pseudoscalar $b \bar b$ bound state
$\eta_b$ is still elusive though experimental search for this state
at the di-gamma decay channel has been initiated recently
\cite{1997}. \\

As an attempt to improve the theoretical predictions involving the phenomenological
description of the meson, using the redial wave functions and other model parameters of
the different potential models we study the decay of $^1 S_0$ quarkonium into di-gamma
and the decay of $^3 S_1$ into lepton pairs using the NRQCD formalism
\cite{BodwinandLepage1995-97}. It is expected that the NRQCD formalism has all the
corrective contributions for the right predictions of the decay rates. NRQCD
factorization expressions for the decay rates of quarkonium and decay are given by
\cite{BodwinGTandPetrelliA2002}

 \beq \label{eq:nq1}
\Gamma(^1 S_0 \rightarrow \gamma \gamma)&=&\frac{F_{\gamma \gamma}(^1S_0)}{m^2_Q} \left
|<0|\chi^{\dag}\psi|^1 S_0>\right|^2 + \frac{G_{\gamma \gamma}(^1S_0)}{m^4_Q} Re \left
[<^1
S_0|\psi^{\dag}\chi|0><0|\chi^{\dag}(-\frac{i}{2}\overrightarrow{D})^2\psi|^1S_0>\right]
\cr && + \frac{H^1_{\gamma \gamma}(^1S_0)}{m^6_Q}<^1
S_0|\psi^{\dag}(-\frac{i}{2}\overrightarrow{D})^2\chi|0><0|\chi^{\dag}(-\frac{i}{2}\overrightarrow{D})^2\psi|^1S_0>\cr
&&+ \frac{H^2_{\gamma \gamma}(^1S_0)}{m^6_Q}\ Re
\left[<^1S_0|\psi^{\dag}\chi|0><0|\chi^{\dag}(-\frac{i}{2}\overrightarrow{D})^4\psi|^1S_0>\right]\eeq

\beq \label{eq:nq2} \Gamma(^3S_1 \rightarrow e^+e^-)&=&\frac{F_{ee}(^3S_1)}{m^2_Q}
\left |<0|\chi^{\dag}\sigma\psi|^3S_1>\right|^2 + \frac{G_{ee}(^3S_1)}{m^4_Q} Re \left
[<^3S_1|\psi^{\dag}\sigma\chi|0><0|\chi^{\dag}\sigma(-\frac{i}{2}\overrightarrow{D})^2\psi|^3S_1>\right]
\cr
&&+\frac{H^1_{ee}(^1S_0)}{m^6_Q}<^3S_1|\psi^{\dag}\sigma(-\frac{i}{2}\overrightarrow{D})^2\chi|0>
<0|\chi^{\dag}\sigma(-\frac{i}{2}\overrightarrow{D})^2\psi|^3S_1>\cr &&+
\frac{H^2_{ee}(^1 S_0)}{m^6_Q}\ Re
\left[<^3S_1|\psi^{\dag}\sigma\chi|0><0|\chi^{\dag}\sigma(-\frac{i}{2}\overrightarrow{D})^4\psi|^3S_1>\right]
\eeq
The short distance coefficients F's and G's of the order of
$\alpha_s^2$ and $\alpha_s^3$ are given by
\cite{BodwinGTandPetrelliA2002}

 \be F_{\gamma \gamma}(^1 S_0)=2
\pi Q^4 \alpha^2 \left[1+\left(\frac{\pi^2}{4}-5 \right) C_F
\frac{\alpha_s}{\pi} \right]\ee

\be G_{\gamma \gamma}(^1S_0)=- \frac{8 \pi Q^4}{3}
 \alpha^2\ee
\be H^1_{\gamma \gamma}(^1S_0)+H^2_{\gamma \gamma}(^1S_0)=\frac{136\pi}{45} Q^4 \alpha^2\ee

\beq F_{ee}(^3S_1) &=& \frac{2 \pi Q^2 \alpha^2}{3}  \{ 1- 4 C_F
\frac{\alpha_s(m)}{\pi}   \cr && +\left[-117.46+0.82n_f+\frac{140
\pi^2}{27} ln(\frac{2m}{\mu_A})\right]\cr &&
(\frac{\alpha_s}{\pi})^2  \} \eeq

\be G_{ee}(^3 S_1)=- \frac{8 \pi Q^2}{9} \alpha^2 \ee
\be H^1_{ee}(^3S_1)+H^2_{ee}(^3S_1)=\frac{58\pi}{54} Q^2 \alpha^2\ee
The matrix elements that contributes to the decay rates of the S wave states into
 $\eta_Q\rightarrow \gamma \gamma$ and $\psi \rightarrow e^+e^-$ through next-to-leading
 order in $v^2$, the vacuum-saturation approximation gives \cite{BodwinandLepage1995-97}

\be
<^1S_0|{\cal{O}}(^1S_0)|^1S_0>=\left|<0|\chi^{\dag}\psi|^1S_0>\right|^2[1+
O(v^4 \Gamma)]\ee

\be
<^3S_1|{\cal{O}}(^3S_1)|^3S_1>=\left|<0|\chi^{\dag}\sigma\psi|^3S_1>\right|^2[1+
O(v^4 \Gamma)]\ee

\beq <^1S_0|{\cal{P}}_1(^1S_0)|^1S_0>= Re
[<^1S_0|\psi^{\dag}\chi|0>&& \cr <0|\chi^{\dag}
(-\frac{i}{2}\overrightarrow{D})^2\psi|^1S_0>] + O(v^4 \Gamma ) \eeq

\beq <^3S_1|{\cal{P}}_1(^3S_1)|^3S_1>=
Re[<^3S_1|\psi^{\dag}\sigma\chi|0> &&\cr
<0|\chi^{\dag}\sigma(-\frac{i}{2}\overrightarrow{D})^2\psi|^3S_1>] +
O(v^4 \Gamma ) \eeq

\beq <^1S_0|{\cal{Q}}^1_1(^1S_0)|^1S_0>=
<0|\chi^{\dag}(-\frac{i}{2}\overrightarrow{D})^2\psi|^1S_0> \eeq

\beq <^3S_1|{\cal{Q}}^1_1(^3S_1)|^3S_1>= <0|\chi^{\dag} \sigma
(-\frac{i}{2}\overrightarrow{D})^2\psi|^3S_1> \eeq

\beq <^1S_0|{\cal{Q}}^2_1(^1S_0)|^1S_0>=
<0|\chi^{\dag}(-\frac{i}{2}\overrightarrow{D})^4\psi|^1S_0>\eeq
\beq <^3S_1|{\cal{Q}}^2_1(^3S_1)|^3S_1>=
<0|\chi^{\dag}\sigma(-\frac{i}{2}\overrightarrow{D})^4\psi|^3S_1>\eeq
The Vacuum saturation allows the matrix elements of some four fermion operators to
 be expressed in terms of the regularized wave-function parameters given by  \cite{BodwinandLepage1995-97}
\be <^1S_0|{\cal{O}}(^1S_0)|^1S_0>=\frac{3}{2 \pi}|R_{P}(0)|^2\ee
\be <^3S_1|{\cal{O}}(^3S_1)|^3S_1>=\frac{3}{2 \pi}|R_{V}(0)|^2\ee
\be <^1S_0|{\cal{P}}_1(^1 S_0)|^1S_0>
 = -\frac{3}{2 \pi}|\overline{R^*_{P}}\
\overline{\bigtriangledown^2 R_{P}}| \ee

\be <^3S_1|{\cal{P}}_1(^3 S_1)|^3S_1>=-\frac{3}{2
\pi}|\overline{R^*_{V}}\ \overline{\bigtriangledown^2 R_{V}}| \ee

\beq <^1S_0|{\cal{Q}}^1_1(^1S_0)|^1S_0>= -\sqrt{\frac{3}{2\pi}}
\overline{\nabla^2} R_{P}\eeq

\beq <^3S_1|{\cal{Q}}^1_1(^3S_1)|^3S_1>=- \sqrt{\frac{3}{2\pi}}
\overline{ \nabla^2} R_{V} \eeq

\beq \label{eq45}
<^1S_0|{\cal{Q}}^2_1(^1S_0)|^1S_0>=\frac{3}{2\pi}
\nabla^2 (\overline{\nabla^2} R_{P})
\eeq
\beq \label{eq46}
<^3S_1|{\cal{Q}}^2_1(^3S_1)|^3S_1>=\frac{3}{2\pi}\nabla^2 (
\overline{ \nabla^2} R_{V})
\eeq

We have computed the $\overline{\nabla}^2 R_{p/v}$ term as per ref \cite{Hafsakhan1996}. Accordingly,
\be \nabla^2 R =\epsilon_B\ R \frac{M}{2}, \ \ \ as \ \ r\rightarrow 0
\ee
 where $\epsilon_B$ is the binding energy and M is the mass of the respective mesonic state. The binding energy is computed as $\epsilon_B=M-(2 m_Q)$. The RHS of the Eqn(\ref{eq45}) and (\ref{eq46}) are computed by assuming that $<p^2>^2\approx<p^4>$.
For comparison, we also compute the decay widths with the conventional V-W formula with
and without the radiative corrections.\\

\begin{table*}
\begin{center}
\caption{Decay rates (in keV) of $0^{-+}$ $\rightarrow$ $\gamma \ \gamma $ and the
relevant correction terms of $\eta_c$ and $\eta_b$ mesons.} \label{tab:34}
\begin{tabular}{clcccccccc}
\m{10}{c}{}\\
\hline \hline Systems & Models& \m{1}{c}{$\Gamma_0$}&  \m{1}{c}{$\Gamma_R$} & \m{1}{c}{$\Gamma$} & \m{2}{c}{$\underline{\Gamma_{NRQCD}}$}&\m{1}{c}{$\Gamma_{NRQCD_{frs}}$}& \m{1}{c}{$\Gamma_{Others}$}\\
&&&&&upto $O(v^2)$&upto $O(v^4$)&&& \\
\hline
&   ERHM    &   7.460   &   -2.855  &   4.605  &4.005 &   4.225   &   --&    \\
    &   BT  &   10.870  &   -4.206  &   6.664   &   6.555&  6.561   & --& 7.2$\pm$0.7$\pm$2.0\cite{ParticleDataGroup2006}      \\
    &   PL(Martin)  &   13.406  &   -6.196  &   7.210  & 8.434&  10.691 &--& 7.500\cite{BraatenErich2003}      \\
$\eta_{c}$  &   Log &   10.937  &   -4.349  &   6.588    & 6.691 & 6.697  &-- &   \\
    &   Cornell &   19.512  &   -6.581  &   12.931  & 13.779 &17.447  &--&9.02 $\pm$ 0.8 \cite{BodwinandLepage1995-97}      \\
    &   CPP$_{\nu}=$0.5 &   8.173   &   -2.635  &   5.538 &2.511  &  6.078 &2.992  &   \\
    &    \ \ \ \ \ \  0.7   &   10.918  &   -3.521  &   7.397 &3.706  &7.810&   4.087&    \\
    &    \ \ \ \ \ \  0.9   &   13.465  &   -4.342  &   9.123 &4.925  &9.391&   5.102 &       \\
    &   \ \ \ \ \ \  1.0    &   14.649  &   -4.724  &   9.925&5.552   &10.077&   5.549 &       \\
    &   \ \ \ \ \ \  1.1    &   15.812  &   -5.099  &   10.713&6.171  &10.761&   6.012 &       \\
    &   \ \ \ \ \ \  1.3    &   17.987  &   -5.800  &   12.187&7.387  & 12.016&  7.095  &       \\
    &   \ \ \ \ \ \  1.5    &   19.971  &   -6.440  &   13.531&8.556  & 13.142&  7.536  &       \\
&   \ \ \ \ \ \  1.7    &    22.788 &-7.026   &15.762&9.700    &14.166&8.170     &       \\
    &   \ \ \ \ \ \  1.9    &   23.502   & -7.578   & 16.924&10.789  &15.139 &8.736 &     \\
    &   \ \ \ \ \ \  2.0    &   24.297  &   -7.835  &   16.462&11.295  &15.596&   9.006  &       \\
    \hline
    &   ERHM    &   0.444   &   -0.114  &   0.326   &0.315&   0.317   &--  &     \\
    &   BT  &   0.574   &   -0.149  &   0.424   &   0.445  & 0.455   &   -- &   \\
    &   PL(Martin)  &   0.406   &   -0.118  &   0.288  &0.312 &  0.340   &--&   0.364 \cite{BodwinandLepage1995-97}   \\
$\eta_{b}$  &   Log &   0.435   &   -0.115  &   0.320  & 0.337 & 0.345   &--&   0.490 \cite{BraatenErich2003}  \\
    &   Cornell &   1.244   &   -0.290  &   0.954 &1.015  &  1.112   &   -- &   \\
    &   CPP$_{\nu}=$0.5 &   0.345   &   -0.086  &   0.259 &0.254  &0.254  & 0.195 &       \\
    &    \ \ \ \ \ \  0.7   &   0.422   &   -0.106  &   0.316&0.310   & 0.311&  0.256  &       \\
    &    \ \ \ \ \ \  0.9   &   0.495   &   -0.124  &   0.371&0.365   &0.366&   0.321   &       \\
    &   \ \ \ \ \ \  1.0    &   0.529   &   -0.132  &   0.397&0.390   &0.391&   0.353   &       \\
    &   \ \ \ \ \ \  1.1    &   0.563   &   -0.141  &   0.422&0.416   & 0.416&  0.386  &       \\
    &   \ \ \ \ \ \  1.3    &   0.626   &   -0.156  &   0.470&0.462   & 0.463&  0.435   &       \\
    &   \ \ \ \ \ \  1.5    &   0.683   &   -0.171  &   0.512&0.505   & 0.505&  0.510  &       \\
&   \ \ \ \ \ \  1.7    &   0.735   &   -0.184  &   0.551 &0.544  & 0.545&  0.570   &       \\
    &   \ \ \ \ \ \  1.9    &  0.786   &   -0.196  &   0.590 &0.582  &0.582&   0.628  &       \\
    &   \ \ \ \ \ \  2.0    &   0.811   &   -0.203  &   0.608&0.600  & 0.601&  0.657  &       \\
\hline\hline
\end{tabular}
\end{center}
ERHM \cite{JNPandya2001,Vinodkumar1999}, BT
\cite{BuchmullerTye1981}, PL (Martin) \cite{Martin1980}, Log
 \cite{Quiggrosner1977}, Cornell \cite{Eichten1978} \\
\end{table*}

\begin{table*}
\begin{center}
\caption{Decay rates (in keV) of $1^{--}$ $\rightarrow$ $l^+ \ l^- $ and the relevant
correction terms of $J/\psi$ and $\Upsilon$ mesons.}\label{tab:36}
\begin{tabular}{cclrccccc}
\m{8}{c}{}\\
\hline \hline Systems & Models &  $\Gamma_{VW}$ & $\Gamma_{rad}$ &
$\Gamma$ &  \m{2}{c}{$\underline{\Gamma_{NRQCD}}$}&  \m{1}{c}{$\Gamma_{NRQCD_{frs}}$}&$\Gamma_{EXP}$\cite{ParticleDataGroup2006}\\
&&&&&upto $O(v^2)$&upto $O(v^4$)&& \\
\hline
&ERHM &5.595&-3.381&2.214&2.543&3.246&--&\\
$J/\psi$&BT&8.152&-4.982&3.170&2.539&2.809&--&5.55$\pm$0.14$\pm$ 0.02\\
&PL (Martin)&10.055&-7.341&2.714&3.311&4.698&--&\\
&Log &8.203&-0.171&3.057&1.967&2.094&--&\\
&Cornell&14.634&-7.701&6.933&7.920&10.294&--&\\
&$CPP_{\nu}$=\ 0.5&6.130&-0.624&5.506&4.212&4.973&0.973&\\
&\ \ \ \ \ \ 0.7&8.189&-0.845&7.344&6.199&7.701&1.676&\\
&\ \ \ \ \ \ 0.9&10.153&-1.065&9.088&8.320&10.815&2.447&\\
&\ \ \ \ \ \ 1.0&11.053&-1.165&9.888&9.353&12.398&2.822&\\
&\ \ \ \ \ \ 1.1&11.946&-1.268&10.678&10.430&14.089&3.227&\\
&\ \ \ \ \ \ 1.3&13.621&-1.463&12.158&12.558&17.550&3.996&\\
&\ \ \ \ \ \ 1.5&15.165&-1.645&13.520&12.605&19.037&4.749&\\
&\ \ \ \ \ \ 1.7&16.582&-1.813&14.769&16.643&24.587&5.467&\\
&\ \ \ \ \ \ 1.9&17.920&-1.982&15.938&18.659&28.232&6.185&\\
&\ \ \ \ \ \ 2.0&18.549&-2.061&16.488&19.634&30.032&6.534&\\
\hline
&ERHM&1.320&-0.540&1.303&1.221&1.228&--&\\
$\Upsilon$&B.T.&1.720&-0.076&1.644&1.249&1.267&--&\\
&PL(Martin)&1.218&-0.761&0.457&0.693&0.774&--&1.340\\
&Log &1.305&-0.032&1.273&0.924&0.943&--&$\pm$0.018\\
&Cornell&3.733&-0.232&3.501&2.025&2.270&--&\\
&$CPP_{\nu}$=\ 0.5&1.035&-0.010&1.025&0.935&0.938&0.710&1.43\cite{agray2005}\\
&\ \ \ \ \ \ 0.7&1.266&-0.013&1.253&1.144&1.148&0.933&\\
&\ \ \ \ \ \ 0.9 &1.485&-0.015&1.470&1.344&1.349&1.165&\\
&\ \ \ \ \ \ 1.0 &1.587&-0.017&1.570&1.436&1.442&1.279&\\
&\ \ \ \ \ \ 1.1&1.690&-0.018&1.678&1.529&1.535&1.397&\\
&\ \ \ \ \ \ 1.3&1.878&-0.020&1.858&1.702&1.709&1.575&\\
&\ \ \ \ \ \ 1.5&2.047&-0.022&2.025&1.857&1.865&1.844&\\
&\ \ \ \ \ \ 1.7&2.206&-0.024&2.182&2.002&2.010&2.057&\\
&\ \ \ \ \ \ 1.9&2.357&-0.026&2.331&2.141&2.149&2.266&\\
&\ \ \ \ \ \ 2.0&2.433&-0.026&2.407&2.210&2.219&2.371&\\
\hline
\end{tabular}
\end{center}
ERHM \cite{JNPandya2001,Vinodkumar1999}, BT
\cite{BuchmullerTye1981}, PL (Martin) \cite{Martin1980}, Log
 \cite{Quiggrosner1977}, Cornell \cite{Eichten1978}\\
\end{table*}

Accordingly, the two photon decay width of the pseudoscalar meson
is given by \cite{AKRai2005}

\be\Gamma({0^{-+}{\rightarrow}{2\gamma}})= \Gamma _0 + \Gamma_R \ee

Here $\Gamma _0$ is the conventional Van Royen-Weisskopf term for the  $0^{-+}
\rightarrow \gamma \gamma$ decays \cite{Vanroyenaweissskopf}, where  $\Gamma_R$ is  due
to the radiative corrections for this decay which is given by

\be \label{eq:0gamma}
 \Gamma_0 = \frac{12 \alpha_e^2 e_Q^4}{M_P^2} \ R_{P}^{2}(0) \ee
and
 \be \Gamma_R=\ \frac{\alpha_s}{\pi} \left(\frac{\pi^2-20}{3}\right) \ \Gamma_0\ee \\

Similarly, the leptonic decay width of the vector meson is computed as
 \be \Gamma({1^{--}{\rightarrow}{l^+ l^-}})=  \Gamma
_{VW}  + \Gamma_{rad}  \ee

where
 \be
\label{eq:vwgamma} \Gamma _{VW}= \frac {4 \alpha_e^2 e_Q^2}{M_V^2} \ R_{V}^{2}(0) \ee
$\Gamma_{rad}$, the radiative correction is given by

\be \Gamma_{rad}=\ \frac{-16}{3 \pi} \alpha_s \ \Gamma _{VW}\ee

It is obvious to note that the computations of the decay rates and
the radiative correction term described here require the right
description of the meson state through its radial wave function at
the origin, R(0) and its mass $M$ along with other model parameters
like $\alpha_s$ and the model quark masses. Generally, due to lack
of exact solutions for colour dynamics, R$_{P/V}$(0) and M are
considered as free parameters of the theory \cite{Hafsakhan1996}.
However, it is appropriate to employ the phenomenological model
spectroscopic parameters such as of the predicted mesonic mass and
the corresponding wave function for the computations of the decay
widths. In many cases of potential model predictions, the radial
wave function at the origin are over estimated as for the decay
rates are concerned. In such cases, it is argued that the decay of
$Q\bar Q$ occurs not at zero separation, but at some finite $Q-\bar
Q$ radial separation. Then arbitrary scaling of the radial wave
function at zero separation are done to estimate the decay rates
correctly \cite{Eichten1978}. In the present computation of the
decay rates using the NRQCD formalism we present our results
obtained by using the radial wave function and their derivatives at
zero separation ($\Gamma_{NRQCD}$) as well as at a finite radial separation of $r_o$, ($\Gamma_{NRQCD_{frs}}$).
We defined $r_o$ by \be r_o=\frac{N_c|e_Q|}{M_{P/V}}\ee of the
mesonic state. It is similar to the compton radius and we call it as
color compton radius of the $Q \bar Q$ systems. Here, $N_{c}=3$ and
$e_{Q}$ is the charge of the quark in terms of the
electron charge.\\

The computed decay widths for $0^{-+}$ $\rightarrow $ $\gamma \
\gamma $, are presented in Table \ref{tab:34} and for $1^{--}$
$\rightarrow$ $l^+ \ l^- $ are listed in Table \ref{tab:36}. In the case of $\Gamma_{NRQCD}$ terms up to $O(v^2)$ and terms up to $O(v^4)$ are separately tabulated to highlight their contributions in the respective decays.

\section{Conclusion and discussion}
In this paper, we have made a comprehensive study of the heavy-heavy
flavour mesonic systems in the general frame work of potential
models. The potential model parameters and the masses of the charmed
and beauty quark obtained from the respective quarkonia mass
predictions have been employed to study their decay properties in
the frame work of NRQCD formalism as well as using the conventional
Van-Royen-Weisskopf nonrelativistic formula. We have also made
parameter free prediction of the weak decay properties of $B_c$
meson. The weak decay constants of the pseudoscalar ($f_P$) and the
vector meson ($f_V$) computed here are is found to be in accordance
with the recent predictions based on relativistic Bethe- Salpeter
method \cite{glwang206}. The departure from the predicted linear
dependence of $f_P$ with the mesonic masses within the effective
light-front model in the heavy flavour sector suggest the
requirement of more refined mechanism related to their wavefunctions
incorporating the confinement and
hyperfine splitting.\\

Masses of the pseudoscalar and vector mesons and the values of the
radial wave function at the origin for $c\bar c$, $c\bar b$ and
$b\bar b$ systems are computed in different potential schemes. The
respective decay constants ($f_P,\ f_V$) are computed with and
without QCD corrections. Using the predicted masses and redial
wave functions at the origin, the digamma, leptonic, light
hadronic decays of quarkonia and the weak decay properties of
$B^+_c$ mesons are studied. For the mass predictions and for the
decay rates the present results based on ($CPP_{\nu}$) are found to be in
accordance with other potential model predictions as well as
with the experimental values.\\

The theoretical ($CPP_{\nu}$) predictions of the decay  widths for
$J/\psi$ $\ra$ $l^+ l^-$ and $\Upsilon$ $\ra$ $l^+ l^-$ as presented
in Table \ref{tab:36} are found to be in accordance with other
potential model predictions with the radiative correction as
well as with the widths computed using NRQCD formalism.\\
Though the radiative corrections are found to be important in most
of the phenomenological models, the NRQCD predictions with their
matrix elements computed at finite radial separation defined through
the 'color compton radius'
 are found to be in better agreement with the experimental values for most of the cases. \\

It is interesting to note that the ERHM\cite{Vinodkumar1999}
predictions of the di-gamma decay widths of $\eta_c$ and leptonic
decay widths of $J/\psi$ and $\Upsilon$ are in good agreement with
the respective experimental results with out any correction to the
Van-Royen-Weisskopf formula.\\

The NRQCD width for $\eta_{c} \ra \gamma \gamma$ predicted in the
present study based on the potential model parameters of BT
\cite{BuchmullerTye1981}, Log \cite{Quiggrosner1977},
CPP$_{\nu}$=0.7,0.9 are close to the experimental value of
7.2$\pm$0.7$\pm$2.0 $keV$ reported by PDG2006
\cite{ParticleDataGroup2006}. However for the $\eta_{b} \ra \gamma
\gamma$ case, most of the model predictions based on NRQCD formalism
are very close to similar theocratical predictions of
\cite{BodwinandLepage1995-97}. The predictions based on V-W formula
with radiative corrections are also found to be in close agreement
with the prediction of\cite{BodwinandLepage1995-97} and \cite{BraatenErich2003} respectively.\\

The predictions of $\eta_{b}$ mass spectra, its hyperfine mass split
 $(\Upsilon -\eta_b)$ of 60 Me$V$, its decay constant
$f_P$ and the digamma width etc are important for the
experimental hunting of $\eta_b$ state.\\

In the case of the dileptonic width of $c\bar{c}$ state, our
predictions based on the NRQCD formalism with the finite range
correction for the inter quark potential index $1.5\leq\nu\leq1.7$
are in fair agrement with the experimental value of 5.55$\pm$0.14
ke$V$; while that $b \bar b$ system  the NRQCD$_{frs}$ prediction is
in good agrement with value of 1.340 $\pm$ 0.018 keV for the
potential index $\nu=1.1$. The CPP$_{\nu=0.5}$ predictions based on
V-W with radiative correction is also found to be in good agreement
with the expected values while in all other choices of $\nu$ over
estimates the decay width. It indicates the importance of the
computation from of the
decay width at finite range of quark-antiquark separation.\\
In the case of the leptonic decay width of $\Upsilon(1S)$ state,
most of the models do provide the decay widths in close agreement
with the expected
value either using NRQCD formalism or using V-W with radiative corrections.\\
Here, again the ERHM prediction for both $J/\psi$ and $\Upsilon$ are
found to be very close to the respective experimental values with
the conventional V-W formula only. It is suggests the adequacy of
the model parameters that provide
the spectroscopy as well as the decay properties.\\
To summarize, we find that the spectroscopy of $c \bar c$ system (1S
to 3S) studied here are in good agreement with the respective
experimental values in the potential range of $1.1\leq \nu \leq1.3$.
However, the spectroscopic predictions with potential index
$\nu=1.5$ for $b \bar b$ system are found to be in agreement with
the respective experimental value. The spectroscopic predictions of
the $b \bar c$ system in the potential range $1.1\leq \nu \leq1.3$
are found to be in accordance with other model
predictions.\\
In the case of the di-gamma decay widths of $c \bar c$ system,
better agreement occurs for the potential index $\nu=0.7$ under the
NRQCD and conventional V-W formula with radiative correction.
However, the NRQCD$_{frs}$ provide the experimental value of the
decay width in the potential index range of $1.3\leq \nu \leq1.5$
only. For the $b \bar b $ system, better consistency in the
predictions of both the leptonic and di-gamma widths are observed
around the potential index $0.7\leq \nu \leq1.1$.

The present study of the decay rates of quarkonia clearly indicates
the relative importance of QCD related corrections on the
phenomenological potential models. The success of potential models
in the determination of the $S$ and $P$ wave masses and decay rates
of $c\bar c$ and $b \bar  c$ and $b\bar b$ systems provide future
scopes to study various transition rate and excited states of these
mesonic systems. With the masses and wave functions of the heavy
flavour mesons at hand, it would be rather simple to compute various
transition rates such as $E1$ and $M1$ in these mesons. Such
computations largely form the future applications of the present
study. The decay rates and branching ratios of heavy flavour mesonic
bound states are important ingredients in
our present understanding of QCD.\\

The semileptonic decays offer an extremely favorable testing
ground for both perturbative QCD, radiative corrections and
nonperturbative QCD effects such as decay constants, form factors,
and the best possible estimations of the CKM matrix elements. With
the mass parameters of the beauty and the charm quark fixed from
the study of its spectra, we have successfully computed the
semi-leptonic decay width of $B_c$-meson.\\

The partial widths obtained here within the spectator model are
compared with those obtained though the Bethe-Salpeter approach
\cite{EL-hady1999} as well as that from a relativistic quark model
\cite{sg} in Table \ref{ta:weakdcy}. We obtained a higher
branching ratio in the $b$-decay channel compared to other
approaches as seen from in Table \ref{ta:weakdcy}. We get about
$64 \%$ as the branching fractions of b-quark decay, about $33 \%$
as that of $c$-quark decay and about $3 \%$ in the annihilation
channel. However, the CKM mixing matrix elements $V_{cb}$ and
$V_{cs}$ used as free parameters in all the three models are
different but lie within the range given in particle data group
\cite{particle2002}. The lifetime of $B^+_{c}$ predicted by the
present calculation is found to be in good agreement  with the
experimental values as well as that by the Bethe Salpeter method
\cite{EL-hady1999}. The predicted values from relativistic model
\cite{sg} is found to be far from the experimental values as well
as other theoretical models.\\
Another aspect of the present study is that the decay of $Q \bar Q $
system occurs at a finite range of its seperation provided by the
color compton radius. This enable us to understand at least
qualitatively the importance of various processes that occur at different radial seperation. \\

In conclusion, we have studied the importance of the spectroscopic
parameters of different potential models in the predictions of the
low-lying states of $c \bar c$, $c \bar b$ and $b \bar b$ systems as
well as their decay properties in the frame work of NRQCD
formalism.\\

{\bf Acknowledgement:} The work is partially supported by Department
of Science and Technology, Government of India under a Major
research project SR/S2/HEP-20/2006. Ajay Kumar Rai Would like to thanks
Namit Mahajan of Physical research Laboratory, Ahmedabad for useful discussion and suggestions. \\


\begin{thebibliography}{99}
\bibitem{ParticleDataGroup2006} Particle Data Group, W. M. Yao, et al.,J.Phys. {\bf G 33}, 1 (2006).
\bibitem{particle2002} Particle Data Group, K. Hagiwara et. al. Phys. Rev. {\bf D 66}, 010001 (2002).
\bibitem{M.B.Voloshin2008}M. B. Voloshin; hep-ph/0711.4556v3 (2008)
\bibitem{Eichten2008}E. Eichten, S. Godfrey, Hanna Mahlke and Jonathan L. Rosner;  hep-ph/0701208v3 (2008)
\bibitem{BuchmullerTye1981} Buchmuller and Tye, Phys.Rev {\bf D24}, 132 (1981).
\bibitem{Martin1980} A. Martin, Phys. Lett. {\bf 93 B } 338 (1980).
\bibitem{Amartin1979} A. Martin, Phys. Lett. {\bf 82 B } 272 (1979).
\bibitem{Richadson1979} J. L. Richardson, Phys. Lett. {\bf 82 B}, 272 (1979).
\bibitem{Quiggrosner1977} C. Quigg and J. L. Rosner, Phys. Lett. {71 B} 153 (1977).
\bibitem{QuiggRosner1979} C. Quigg and J. L. Rosner, Phys. Rept. {\bf 56} 167 (1979).
\bibitem{Eichten1978}E. Eichten and  K. Gottfried, T. Kinoshita, K. D. Lane and T. M. Yan, Phys. Rev. {\bf D 17}, 3090(1978).
\bibitem{vijayakumar2004} Vijaya Kumar K B, B Hanumaiah and S Pepin, Eur.Phys. J. {\bf A 19} 247 (2004).
\bibitem{Altarelli1982} Altarelli G, Cabibbo N, Corbo G, Maiani L and Martinelli G, Nucl. Phys. {\bf B 208.} 365 (1982).
\bibitem{Ebert2003} D. Ebert R.N. Faustov and V. O. Galkin, Phys. Rev. {\bf D 67} 014027 (2003).
\bibitem{SNGupta1996} S. N. Gupta, J. M. Johnson and W. W. Repko, Phys. Rev. {\bf D 54}, 2075 (1996).
\bibitem{SGogfrey1986} S. Godfrey, Phys. Rev. {\bf D 33}, 1391 (1986).
\bibitem{Khadkikar1983} S. B. Khadkikar and S. K. Gupta, Phys. Lett. {\bf B 124}, 523(1983).
\bibitem{hawng1996} Hwang D S, Kim C S and Wuk Namgung, Phys. Rev. {\bf D 53}, 4951  (1996).
\bibitem{JNPandya2001} J N Pandya and P C Vinodkumar, Pramana J. Phys {\bf 57}, 821 (2001).
\bibitem{Vinodkumar1999} P. C. Vinodkumar, J. N. Pandya, V. M. Bannur and S. B. Khadkikar, Eur. Phys. J. {\bf A 4}, 83 (1999).
\bibitem{AKRai2002} Ajay Kumar Rai, R H Parmar and P C Vinodkumar Jnl. Phys G. {\bf 28}, 2275 (2002).
\bibitem{AKRai2005} Ajay Kumar Rai, J N Pandya and P C Vinodkumar Jnl. Phys G. {\bf 31}, 1453 (2005).
\bibitem{AKRai2006} Ajay Kumar Rai and P C Vinodkumar, Pramana J. Phys. {\bf 66}, 953 (2006).
\bibitem{EL-hady1999} A. Adb El-Hady, M.A.K. Lodhi and J. P. Vary, Phys. Rev. {\bf D 59} 094001 (1999).
\bibitem{N barmbilla2005}N. Barmbilla, et al., CERN Yellow Report, CERN-2005-005, Geneva: arXive:hep-ph/0412158.
\bibitem{AppelquistTandPolitzerHD1975} Appelquist T and Politzer H D, Phys. Rev. Lett. {\bf 34}, 43 (1975).
\bibitem{BarbieriRandGatto1976} Barbieri R, Gatto  R and Kogerler R,Phys. Lett. {\bf 60 B}, 183 (1976).
\bibitem{BarbieriR1979} Barbieri R et al, Nucl. Phys. {\bf B 154}, 535 (1979).
\bibitem{HagiwaraKandKim1981} Hagiwara K, Kim C B and T Yosino, Nucl. Phys. {\bf B 177}, 461 (1981).
\bibitem{MeackenziePandLepage1981} Meackenzie P and Lepage G P, Phys. Rev. Lett. {\bf 47}, 1244 (1981).
\bibitem{BodwinGTandPetrelliA2002} Bodwin G T and Petrelli A, Phys. Rev. {\bf D 66} 094011 (2002).
\bibitem{BarbieriRandCaffo1980-1} Barbieri R, Caffo M, Gatto R and Remiddi E, Phys. Lett. {\bf 95 B}, 93 (1980); Nucl. Phys.{\bf B
192} 61 (1981).
\bibitem{BodwinandLepage1995-97} Bodwin G T, Braaten Erich and Lepage G P, Phys. Rev. {\bf D 51}, 1125 (1995); {\bf 55} (E) (1997).
\bibitem{APineda1998}A. Pineda and J. Soto, Nucl. Phys. {\bf B}, Proc. Suppl, {\bf 64} (1998) 428.
\bibitem{NoraBrambilla2003} Nora Brambilla, Dolors Eiras, Antonia Pineda and Joan Soto, Antonio Vairo, Phys. Rev. {\bf D 67} 034018 (2003).
\bibitem{NoraBrambilla2005} Nora Brambilla, Antonia Pineda and Joan Soto, Antonio Vairo, Rev. Mod. Phys. {\bf 77} 1423 (2005) (and reference therein)

\bibitem{Chienhep-ph/0609036} Chin-Wen Hwang and Zheng-Tao Wei,J. Phys. G. {\bf 34} 687 (2007); hep-ph/0609036.
\bibitem{hycheng1997}H. Y. Cheng C. Y. Cheung and C. W. Hwang, Phys. Rev. {\bf D 55}, 1159 (1997).
\bibitem{Sameer2004-5} Sameer M. Ikhdair and Ramazan Sever Int. J. Mod. Phys. {\bf A 19} 1771 (2004) {\it ibid} {\bf A 20} 4035 (2005).
\bibitem{Motyka1998} L. Motyka and K. Zalewiski, Eur. Phys. J. {\bf C 4 } 107 (1998); $ibid$., Z. Phys. {\bf C 69 } 343 (1996).
\bibitem{xtsong1991} X. T. Song, J. Phys. {\bf G 17} 49 (1991).
\bibitem{Sterett1997} Sterrett J. Collins, T.D. Imbo, B. A. King, E. C. Martell, Phys. Lett. {\bf B 393} 155 (1997).
\bibitem{Eichten1994} Eichten E J and  Quigg, Phys. Rev. {\bf D 49}, 5845 (1994).
\bibitem{SNGuPta1982} S. N. Gupta, Stanley  and W. W. Repko, Phys. Rev. D. {\bf 26} 3305 (1982).
\bibitem{Gerstein1995} Gershtein S S, Kiselev V V, Likhoded A K and Tkabladze A V, Phys. Rev. {\bf D 51}, 3613 (1995).
\bibitem{Vanroyenaweissskopf} R. Van Royen and V. F. Weisskopf, Nuovo Cimento {\bf 50}, 617 (1967).
\bibitem{1997} D. S. Hwang and Gwang-Hee Kim, Z. Phys. {\bf C 76} 107 (1997).
\bibitem{EBraaten1995} E. Braaten and S. Fleming, Phys. Rev. {\bf D 52} 181 (1995).
\bibitem{Gershtein1998} S. S. Gershetin, hep-ph/983433 (1998).
\bibitem{Abe2002} Abe  et al, Bell Collaboration, Phy. Rev. Lett. {\bf 89} 42001
(2002).
\bibitem{Saldo2003} L. A. M. Saledo, et al.arxive:hep-ph/0311008.
\bibitem{sg}Stephen Godfrey, Phys. Rev. {\bf D 70}, 054017 (2004) arXiv:hep-ph/0406228 (2004).
\bibitem{CDFcollabration} CDF Collabration, Phys. Rev. {\bf D 58}, 112004 (1998).
\bibitem{sfredford2007}S. F. Redford and W. W. Repko, Phys. Rev. D {\bf 75}, 074031 (2007).
\bibitem{Cleocollaboration2001} K. W. Edwards et al., CLEO Collaboration, Phys. Rev.
Lett. {\bf 86}, 30 (2001).
\bibitem{glwang206} G. L. Wang, Phys Lett {\bf B 633} 492 (2006).
\bibitem{Hafsakhan1996} Hafsakhan and Pervez Hoodbhoy,  Phys. Rev. {\bf D53}, 2534 (1996).
\bibitem{degoom2000} Particle Data Group,D. E. Groom et al., Euro. Phys. J. {\bf C 15} 1 (2000)
\bibitem{BraatenErich2003} Braaten Erich and Lee J, Phys. Rev. {\bf D 67} 054007 (2003).
\bibitem{NFabianoGPa} N. Fabiano and G. Pancheri, Eur. Phys. J. {\bf 25}, 421 (2002).
\bibitem{agray2005} A. Gray, I. Allison, C. T. H. Davies, E. Gulez, G. P. Lepage, J. Shigemitsu and M. Wingate, Phys. Rev. {\bf 72} 094507 (2005).  hep-lat/0507013.
\end{thebibliography}
\end{document}